\documentclass[a4paper,11pt]{article}   
\usepackage{lineno,hyperref}
\usepackage[utf8]{inputenc}
\usepackage[english]{babel}
\usepackage{ifthen}
\usepackage{amssymb,amsmath}
\usepackage[usenames]{color}
\date{}

 \newif\ifNoRemark
    \def\addtheorem#1#2#3#4{ 
    \ifthenelse{\expandafter\isundefined\csname the#2\endcsname}{\newcounter{#2}}{}
    \newenvironment{#1}[1][\global\NoRemarktrue]
     {\par\addvspace{2mm}\noindent
       \refstepcounter{#2}{\bf #3~\csname the#2\endcsname
      \vphantom{##1}\ifNoRemark.\ \else\ (##1).\fi}\begingroup #4}%
     {\endgroup\par\addvspace{1mm}\global\NoRemarkfalse}
    \expandafter\newcommand\csname b#1\endcsname{\begin{#1}}
    \expandafter\newcommand\csname e#1\endcsname{\end{#1}}
    }

\addtheorem{theorem}{thrm}{Theorem}{\sl}
\addtheorem{lemma}{lmm}{Lemma}{\sl}
\addtheorem{proposition}{prpstn}{Proposition}{\sl}
\addtheorem{corollary}{crllr}{Corollary}{\sl}
\addtheorem{problem}{problem}{Problem}{\sl}
\addtheorem{remark}{}{Remark}{}

\begin{document}

\title{An Upper Bound on the Number of Bent  Functions
\thanks{The work was supported by the program of fundamental scientific
researches of the SB RAS  I.5.1, project No. 0314-2019-0017.} }

\author{ Vladimir N. Potapov\\ {\em \small Sobolev Institute of Mathematics, Novosibirsk,
Russia; email: vpotapov@math.nsc.ru}}

\maketitle

\begin{abstract}
 The number of $n$-ary bent
 functions  is less than
$2^{3\cdot2^{n-3}(1+o(1))}$ as $n$ is even and $n\rightarrow\infty$.
\end{abstract}


\section{Introduction}

Let $F= \{0, 1\}$. The set $F^n$ is called the $n$-dimensional
boolean hypercube (or the boolean $n$-cube). The hypercube $F^n$
equipped with scalar multiplication and coordinate-wise modulo 2
addition $\oplus$ can be considered as an $n$-dimensional vector
space. Functions
 $\phi_x(y)=(-1)^{\langle x,y\rangle}$ are called characters. Here $\langle
x,y\rangle=x_1y_1\oplus\dots\oplus x_ny_n$ is the inner product. Let
$f$ be a function that  maps from the boolean hypercube to real
numbers. The Fourier transform of $g$ is defined by the formula
$\widehat{g}(y)=(g,\phi_y)$, i.e., $\widehat{g}(y)$ are the
coefficients of the expansion of $f$ in the basis of characters.

We can define the Walsh--Hadamard transform of boolean function
$f:F^n\rightarrow F$  by the formula $W_f(y)=\widehat{(-1)^b}(y)$,
i.e., $$W_f(y)=\sum\limits_{x\in F^n}(-1)^{f(x)\oplus \langle
x,y\rangle}.$$ Boolean function $b$ is said to be a bent function if
$W_b(y)=\pm 2^{n/2}$ for all $y\in F^n$. It is easy to see that bent
functions $b:F^n\rightarrow F$ exist only if $n$ is even.

Every boolean function $f$ can be represented as a polynomial $
f(x_1,\dots,x_n)=\bigoplus\limits_{y\in
    F^n}M[f](y)x_1^{y_1}\cdots x_n^{y_n},$
     where $x^0=1, x^1=x$, and
$M[f]:F^n\rightarrow F$ is the M\"obius transform of $f$. It is well
known that
\begin{equation}\label{equpperbent0}
M[f](y)=\bigoplus\limits_{x\in \Gamma_y}f(x)
\end{equation}
where $\Gamma_y$ is the face of $F^n$ that contains $\bar 0$, $y$
and all vectors between $\bar 0$ and $y$. Note that $M[M[f]]=f$ for
each boolean function.
 The degree of this polynomial   is called  the algebraic
degree  of $f$. It is easy to see that the number of boolean
functions of $n$ variables with degree ${\rm deg}\,f\leq d$ is equal
to $2^{\sum\limits_{i=0}^d {n \choose i}}$.

It is well known (see \cite{Carlet}--\cite{Mes0}) that the degree of
a bent function $b:F^n\rightarrow F$ is not greater than $n/2$.
Therefore the number of bent functions is not greater than
$2^{\sum\limits_{i=0}^{n/2} {n \choose i}}$. Let $N_n$ be the number
of bent functions on $n$ variables. Then $\log N_n\leq 2^{n-1}+
\frac12{n \choose n/2} $. In \cite{CK} and \cite{Ag} there are some
upper bounds of $N_n$.
 These bounds have type $\log
N_n\leq 2^{n-1}(1+o(1))$ and they are a bit better than the trivial
upper bound based on the estimation of algebraic degree. We  obtain
new upper bound $\log N_n\leq 3\cdot2^{n-3}(1+o(1))$. Note that
Tokareva's conjecture (see \cite{Tok}) of the decomposition of
boolean functions into sums of bent functions implies that $\log
N_n\geq 2^{n-2}+\frac12{n \choose n/2}$.

\section{Preliminaries}

It is well known (see \cite{Carlet},\cite{Tsf}) that
$$\widehat{f*g}= {\widehat{f}\cdot\widehat{ g}} \qquad {\rm
and}\qquad \widehat{(\widehat{f})}=2^nf$$ where
$f*g(z)=\sum\limits_{x\in F^n}f(x)g(z\oplus x)$. Consequently, it
holds
\begin{equation}\label{equpperbent11}
2^{n}f*g= \widehat{\widehat{f}\cdot\widehat{ g}}.
\end{equation}
 Let $\Gamma$, $\bar 0\in \Gamma$, be a face of hypercube and let $\Gamma^\perp$ be the
dual face, i.e., $\Gamma^\perp=\{y\in F^n : \forall x\in \Gamma,
\langle x,y\rangle=0 \}.$ Denote by ${\bf1}_S$ an indicator function
of a set $S$. If $g={\bf1}_{\Gamma^\perp}$ then by
(\ref{equpperbent11}) it follows
\begin{equation}\label{equpperbent1}
{f}*{\bf1}_{\Gamma^\perp}=2^{-\mathrm{dim}\,\Gamma}\widehat{\widehat{f}\cdot{\bf1}_{\Gamma}}
\end{equation}
for any face $\Gamma\subset F^n$, $\bar{0}\in \Gamma$.
\begin{lemma}
Suppose that $f$ and $g$ are boolean functions in $n$ variables. For
any face $\Gamma\subset F^n$ if $W_f|_\Gamma=W_g|_\Gamma$ then
$\sum\limits_{x\in z\oplus\Gamma^\perp}(-1)^{f(x)}=\sum\limits_{x\in
z\oplus\Gamma^\perp}(-1)^{g(x)}$ for any $z\in F^n$.
\end{lemma}
Proof. It follows from (\ref{equpperbent1}). Indeed it holds
$\widehat{(-1)^f}\cdot{\bf1}_{\Gamma}=\widehat{(-1)^g}\cdot{\bf1}_{\Gamma}$
by conditions of the lemma. Then
$(-1)^f*{\bf1}_{\Gamma^\perp}=(-1)^g*{\bf1}_{\Gamma^\perp}$. It is
clear that $((-1)^f*{\bf1}_{\Gamma^\perp})(z)=\sum\limits_{x\in
z\oplus\Gamma^\perp}(-1)^{f(x)}$. Hence we obtain the required
conclusion. $\square$

Denote by $\mathrm{wt}(z)$  the number of units in $z\in F^n$. Let
$B_r$ be a ball with radius $r$ in $F^n$, i.e., $B_r=\{x\in F^n :
\mathrm{wt}(x)\leq r\}$.

\begin{lemma}
Suppose that $f$ and $g$ are $n$-ary boolean functions  and
$\max\{{\rm deg}(f),{\rm deg}(g)\}\leq r$. If $f|_{B_r}=g|_{B_r}$
then $f=g$.
\end{lemma}
Proof.
 By the hypothesis  of the
lemma we have
    $M[f](y)=M[g](y)=0$ if $\mathrm{wt}(y)>r$. By (\ref{equpperbent0})
     for any $y\in F^n$ such that $\mathrm{wt}(y)=r+1$  we obtain
$$M[f](y)=\bigoplus\limits_{x\in \Gamma_y}f(x)=
f(y)\oplus\bigoplus\limits_{x\in \Gamma_y\cap B_r}f(x) =$$
$$=f(y)\oplus\bigoplus\limits_{x\in \Gamma_y\cap
B_r}g(x)=M[g](y)\oplus f(y)\oplus g(y).$$ Therefore $f(y)= g(y)$ for
any $y\in B_{r+1}$. By induction on weights $\mathrm{wt}(y)$ of the
vectors  $y\in F^n$  we obtain that $f(y)= g(y)$ for all $y\in F^n$.
$\square$

We will use the following property of bent functions.

\begin{proposition}{(\cite{Carlet}--\cite{Mes0})}\label{splat12}
Let $f:{F}^n\rightarrow {F}$ be a bent function, let
$A:{F}^n\rightarrow {F}^n$ be a non-degenerate affine transformation
and let $\ell:{F}^n\rightarrow {F}$ be an affine function. Then
$g=f\circ A\oplus\ell$ is a bent function.
\end{proposition}

Functions $f$ and $g$ from Proposition 1 are called equivalent. It
is easy to see that the cardinality of any equivalence class is not
greater than $a_n=2^{n^2}(1+o(1))$.

\section{Main result}

\begin{theorem} The number of bent functions in $n$ variables is not
greater than $6^{3\cdot2^{n-6}}2^{\cdot2^{n-2}(1+o(1))}$ as $n$ is
even and $n\rightarrow\infty$.
\end{theorem}
Proof. Let $b$ be a bent function in $n$ variables. It is well known
(see \cite{Carlet}--\cite{Mes0}) that $W_b=2^{n/2}(-1)^g$ where $g$
is a bent function too. Therefore ${\rm deg}(g)\leq \frac{n}{2}$.
Consider a face $\Gamma$, $\bar 0\in \Gamma$ and $\dim\Gamma=n-2$.
By Lemma 2 there exist at most $T_n=2^{\sum\limits_{i=0}^{n/2}{n-2
\choose i}}= 2^{2^{n-3}(1+o(1))}$ different functions
$g':\Gamma\rightarrow F$ such that $g'=g|_\Gamma$ and $g$ are  bent
functions in $n$ variables. By Lemma 1 we obtain that $g'$ is
determined by sums $\sum\limits_{x\in
z\oplus\Gamma^\perp}(-1)^{b(x)}$ for all $z\in F^n$.  It is easy to
see that the sums $\sum\limits_{x\in
z\oplus\Gamma^\perp}(-1)^{b(x)}$ can be equal to $-4,-2,0,2$ or $4$.
If a sum equals $\pm 4$ then $b(x)=1$  for all $x\in
z\oplus\Gamma^\perp$ or $b(x)=0$ for all $x\in z\oplus\Gamma^\perp$.
If it equals $\pm 2$ then we have four possibilities for the vector
of values of $b$ on ${z\oplus\Gamma^\perp}$. At last if
$\sum\limits_{x\in z\oplus\Gamma^\perp}(-1)^{b(x)}=0$ then we have
six possibilities for $b|_{z\oplus\Gamma^\perp}$.

For any bent function $b$ the proportion of a $2$-dimensional
subspace $S$ of $F^n$ such that $\sum\limits_{x\in S}(-1)^{b(x)}=\pm
2$ is equal to $\frac12$ (for example, see \cite{PA}). Then we can
find a non-degenerate linear transformation $A$ of $b$ such that the
proportion of $2$-dimensional faces $z\oplus\Gamma^\perp$ with
$\sum\limits_{x\in z\oplus\Gamma^\perp}(-1)^{b\circ A(x)}=\pm 2$ is
not greater than $\frac12$.

Let $\mathcal{L}$ be the set of all affine functions. If
$\sum\limits_{x\in z\oplus\Gamma^\perp}(-1)^{b\circ A(x)}=0$ then
$\sum\limits_{x\in z\oplus\Gamma^\perp}(-1)^{b\circ A(x)\oplus
\ell(x)}=\pm 4$ for $\frac14$ of functions $\ell\in \mathcal{ L}$.
Therefore we can find $\ell_0$ with such proportion or better.

At last we note that  by Lemma 2 it is necessary to recover $b$ only
on $B_{n/2}$. Suppose that $b$ is a function with the best
proportion of sums $-4,-2,0,2,4$ from some equivalence class. Denote
by $Q_n$ the minimum number of elements from $\Gamma$ that is needed
for recovering  $b$. It is easy to see that $Q_n={2^{n-3}(1+o(1))}$.
Since $b'(x)=b(x)\oplus a$ is a bent function, without loss of
generality, we suppose that $b$ has the best proportion of sums
$-4,-2,0,2,4$ on faces which is determined by this $Q_n$ elements.
Therefore an upper bound of the number of bent functions is
$a_nT_n4^{Q_n/2}6^{3Q_n/8}$ where $a_n$ is the cardinality of the
largest equivalence class. $\square$

It is easy to see that $6^{3\cdot2^{n-6}}2^{2^{n-2}}$ is less than
$2^{3\cdot2^{n-3}}$.

\end{document}